\documentclass{article}

\usepackage{arxiv}

\usepackage[utf8]{inputenc} 
\usepackage[T1]{fontenc}    
\usepackage{hyperref}       
\usepackage{url}            
\usepackage{booktabs}       
\usepackage{amsfonts}       
\usepackage{nicefrac}       
\usepackage{microtype}      
\usepackage{lipsum}
\usepackage{rotating}
\usepackage{subcaption}
\usepackage{lscape}
\usepackage{graphicx}
    \usepackage{listings}
\graphicspath{ {./images/} }

\title{Comprehensive evaluation of no-reference image quality assessment algorithms
on authentic distortions}

\author{
 Domonkos Varga 
}

\begin{document}
\maketitle
\begin{abstract}
Objective image quality assessment deals with the prediction of digital images' perceptual
quality. No-reference image quality assessment predicts the quality of a given input image
without any knowledge or information about its pristine (distortion free) counterpart.
Machine learning algorithms are heavily used in no-reference image quality assessment
because it is very complicated to model the human visual system's quality perception.
Moreover, no-reference image quality assessment algorithms are evaluated
on publicly available benchmark databases. These databases contain images with their
corresponding quality scores. In this study, we evaluate
several machine learning based NR-IQA methods and one opinion unaware
method on databases consisting of authentic distortions. Specifically, LIVE In the Wild
and KonIQ-10k databases were applied to evaluate the state-of-the-art. For machine learning based
methods, appx. 80$\%$ were used for training and the remaining 20$\%$ were used for testing.
Furthermore, average PLCC, SROCC, and KROCC values were reported over 100 random train-test
splits. The statistics of PLCC, SROCC, and KROCC values were also published using boxplots.
Our evaluation results may be helpful to obtain a clear understanding about the
status of state-of-the-art no-reference image quality assessment methods.
\end{abstract}

\keywords{no-reference image quality assessment}

\section{Introduction}
After digital images are captured, they suffer from a wide variety of distortions
due to compression, editing, retargeting, and transmission. The above mentioned mechanisms
influence the perceived quality of digital images. \textit{Objective image quality assessment
(IQA)} deals with
the measurement of perceptual image quality where image processing, computer vision, artificial
intelligence, and machine learning techniques are applied to construct computational models to
predict perceived image quality. Due to our limited understanding of the human visual system, it is
very complicated to model the human visual system's quality perception. Thus, machine learning
methods are heavily used to imitate the processes of such complex mechanisms of the human visual system.
Since humans are the end users of digital images, the most reliable method to evaluate the quality
of digital images is to ask humans for quality ratings of images. \textit{Subjective IQA} deals with
gathering subjective scores from human observers considering the recommendations of different
international standards, such as
ITU-R BT.500-11 \cite{bt2002methodology} or ITU-P.910 \cite{itu1999subjective}.
Subsequently,
the scores are processed to determine the
mean opinion score (MOS) or the difference mean
opinion score (DMOS). There are several IQA databases publicly available 
which contain images labelled with MOS or DMOS. Table \ref{table:iqadatabase} allows a comparison
of publicly available IQA databases. The majority of IQA data sets contains
a small set of pristine (distortion free) images and distorted images are derived from
the pristine image using different artificial distortion types and levels. On the other hand, some
databases (LIVE In the Wild \cite{ghadiyaram2015massive} 
and KonIQ-10k \cite{lin2018koniq}) consist of individual images with authentic distortions collected
from the Internet or private collections. Objective IQA algorithms are trained and tested on these
databases.

In this study, our goal is to evaluate several state-of-the-art no-reference image quality
assessment (NR-IQA) algorithms, including BLIINDS-II \cite{saad2012blind}, BMPRI \cite{min2018blind},
BRISQUE \cite{mittal2012no}, CurveletQA \cite{liu2014no},
DIIVINE \cite{moorthy2011blind}, ENIQA \cite{chen2019no}, GRAD-LOG-CP \cite{xue2014blind},
GWH-GLBP-BIQA \cite{li2016no}, IQVG \cite{gu2013learning},
MultiGAP-GPR \cite{varga2020multi}, MultiGAP-SVR \cite{varga2020multi},
NBIQA \cite{ou2019novel}, OG-IQA \cite{liu2016blind}, ORACLE \cite{oszust2018optimized},
PIQE \cite{venkatanath2015blind}, SCORER \cite{oszust2019local}, SEER \cite{oszust2019no},
SPF-IQA \cite{varga2020no}, and SSEQ \cite{liu2014noS},
on databases with authentic distortions. Excluding
PIQE \cite{venkatanath2015blind} which is an opinion-unaware method,
all algorithms are machine learning based. Machine learning based algorithms were trained on
appx. 80$\%$ of images and tested on the remaining 20$\%$. Correlation between the predicted
and the ground-truth scores was measured by Pearson's linear correlation coefficient
(PLCC), Spearman's rank order
correlation coefficient (SROCC),
and Kendall's rank order correlation coefficient (KROCC). In this study, we report on the
average PLCC, SROCC, and KROCC values measured over 100 random train-test splits. Moreover,
the statistics of PLCC, SROCC, and KROCC values are summarized with boxplots.

\begin{table*}[ht]
\caption{Comparison of several publicly available IQA databases.
} 
\centering 
\begin{center}
    \begin{tabular}{ |c|c|c|c|c|c|}
    \hline
Database&Ref. images&Test images&Resolution& Distortion levels&Number of distortions\\
    \hline
LIVE \cite{sheikh2006statistical}&29&779&$768\times512$&4-5&5 \\
A57 \cite{chandler2007vsnr}&3&54&$512\times512$&6&3 \\
Toyoma-MICT \cite{horita2011mict}&14&168&$768\times512$&6&2 \\
TID2008 \cite{ponomarenko2009tid2008}&25&1,700&$512\times384$&4&17 \\
CSIQ \cite{larson2010most} & 30& 866&$512\times512$&4-5 &6 \\
VCL-FER \cite{zaric2012vcl} & 23 & 552 &$683\times512$ &6& 4 \\
LIVE Multiple Distorted \cite{jayaraman2012objective}&15 & &$1280\times720$&3& \\
TID2013 \cite{ponomarenko2015image}&25&3,000&$512\times384$&5&24 \\
CID:IQ \cite{liu2014cid}&23&690&$800\times800$&5&6 \\
LIVE In the Wild \cite{ghadiyaram2015massive}&-&1,169&$500\times500$&-&N/A \\
MDID \cite{sun2017mdid}&20&1,600&$512\times384$&4&5 \\
KonIQ-10k \cite{lin2018koniq}&-&10,073&$1024\times768$&-&N/A \\
KADID-10k \cite{lin2019kadid} & 81 &10,125&$512\times384$&5&25\\
 \hline
 \end{tabular}
\end{center}
\label{table:iqadatabase}
\end{table*}

\subsection{Contributions}
The main contribution of this study is the comprehensive evaluation of several
state-of-the-art NR-IQA algorithms
(BLIINDS-II \cite{saad2012blind}, BMPRI \cite{min2018blind},
BRISQUE \cite{mittal2012no}, CurveletQA \cite{liu2014no},
DIIVINE \cite{moorthy2011blind}, ENIQA \cite{chen2019no}, GRAD-LOG-CP \cite{xue2014blind},
GWH-GLBP-BIQA \cite{li2016no}, IQVG \cite{gu2013learning},
MultiGAP-GPR \cite{varga2020multi}, MultiGAP-SVR \cite{varga2020multi},
NBIQA \cite{ou2019novel}, OG-IQA \cite{liu2016blind}, ORACLE \cite{oszust2018optimized},
PIQE \cite{venkatanath2015blind}, SCORER \cite{oszust2019local}, SEER \cite{oszust2019no},
SPF-IQA \cite{varga2020no}, and SSEQ \cite{liu2014noS})
on LIVE In the Wild and KonIQ-10k databases. Average PLCC, SROCC, and KROCC are
reported measured over 100 random train-test splits 
(appx. 80$\%$ for training and 20$\%$ for testing). Moreover, the statistics of
PLCC, SROCC, and KROCC values are summarized in box plots. 

\section{Experimental results}
\label{sec:experimental}
In this section, NR-IQA algorithms are evaluated on LIVE In the Wild and KonIQ-10k databases.
As already mentioned, 80$\%$ of images was used for training and the remaining 20$\%$ for testing.
Furthermore, average PLCC, SROCC, and KROCC values are reported measured over 100 random train-test splits.

\subsection{LIVE In the Wild}
\begin{figure}
    \centering
    \begin{subfigure}[b]{0.45\textwidth}
        \includegraphics[width=\textwidth]{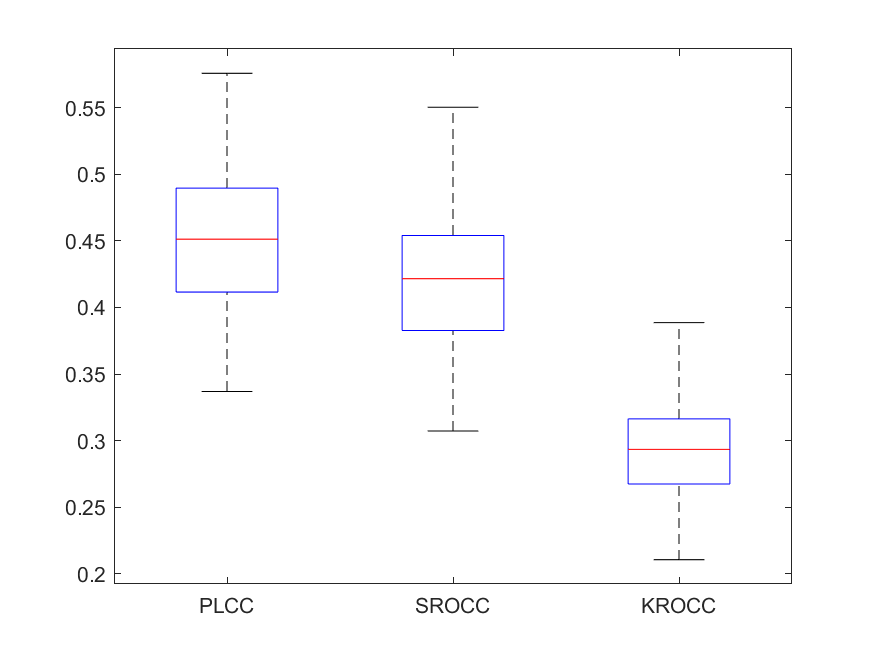}
        \caption{BLIINDS-II.}
    \end{subfigure}
    ~ 
    \begin{subfigure}[b]{0.45\textwidth}
        \includegraphics[width=\textwidth]{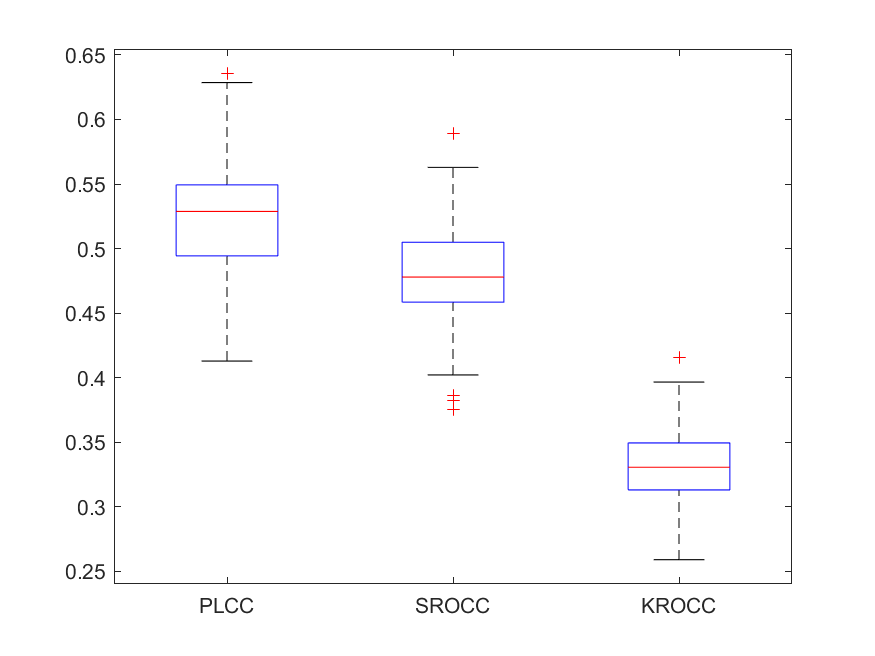}
        \caption{BMPRI.}
    \end{subfigure}

    \quad
    
    \begin{subfigure}[b]{0.45\textwidth}
        \includegraphics[width=\textwidth]{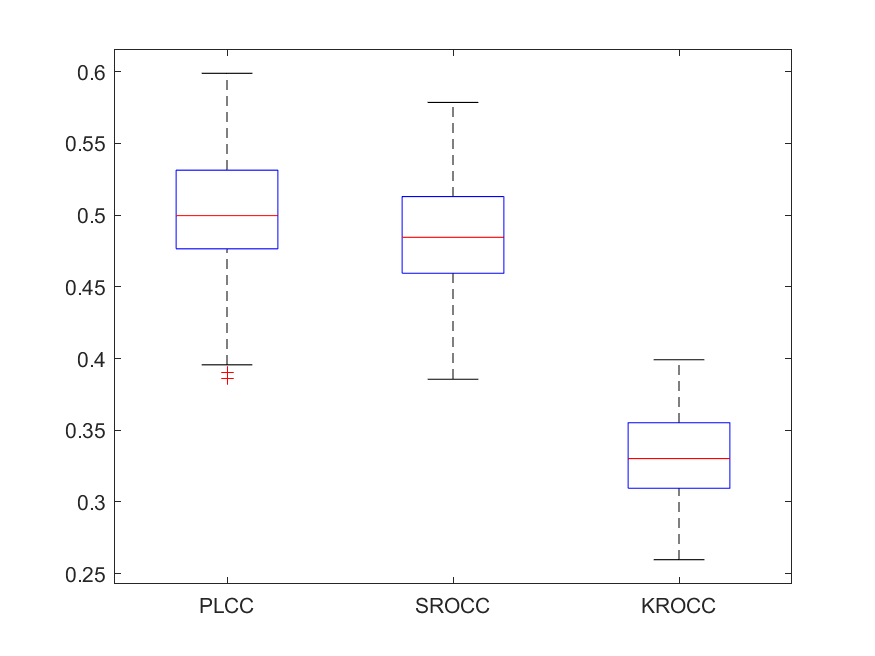}
        \caption{BRISQUE.}
    \end{subfigure}
    ~
    \begin{subfigure}[b]{0.45\textwidth}
        \includegraphics[width=\textwidth]{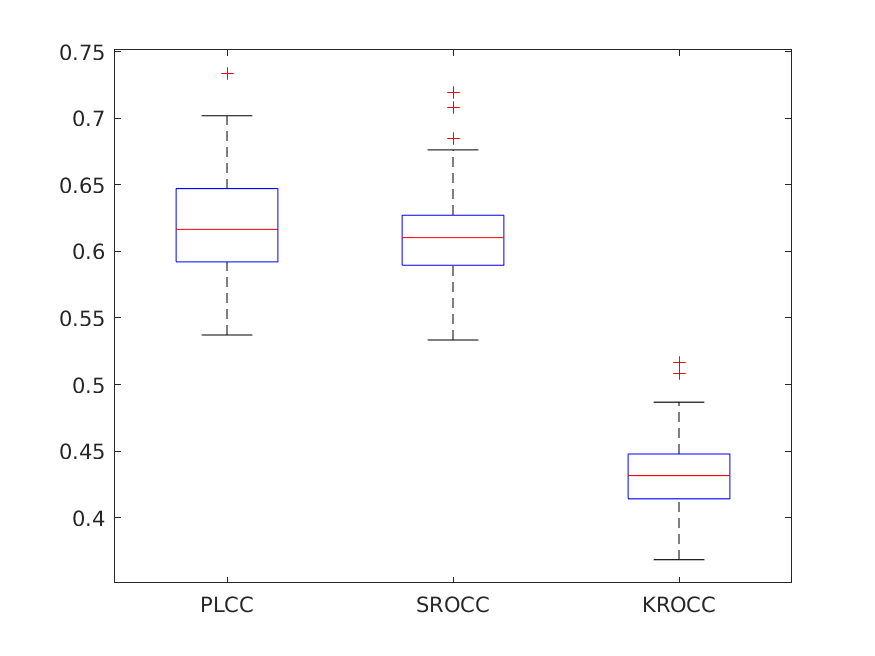}
        \caption{CurveletQA.}
    \end{subfigure}

    \quad

    \begin{subfigure}[b]{0.45\textwidth}
        \includegraphics[width=\textwidth]{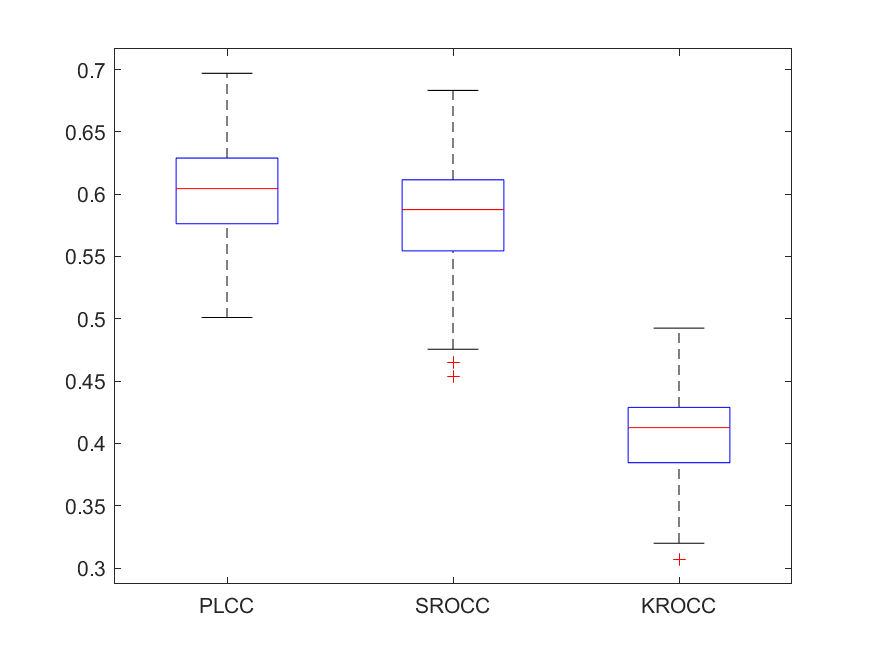}
        \caption{DIIVINE.}
    \end{subfigure}
    ~
    \begin{subfigure}[b]{0.45\textwidth}
        \includegraphics[width=\textwidth]{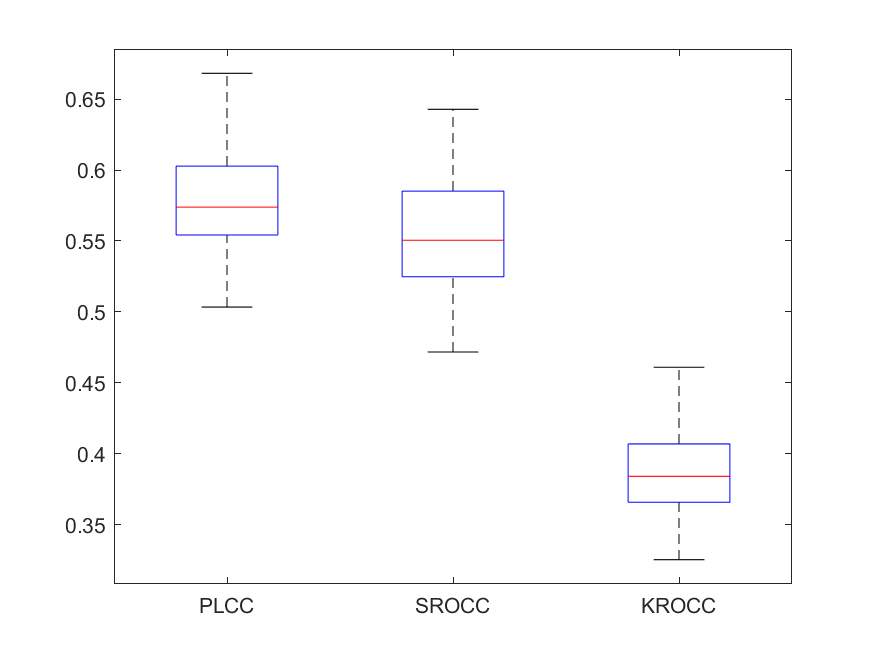}
        \caption{ENIQA.}
    \end{subfigure}
    \caption{Box plots of the measured PLCC, SROCC, and KROCC values on LIVE In the Wild database.
    100 random train-test splits.}
    \label{fig:boxplot1}
\end{figure}

\begin{figure}
    \centering
    \begin{subfigure}[b]{0.45\textwidth}
        \includegraphics[width=\textwidth]{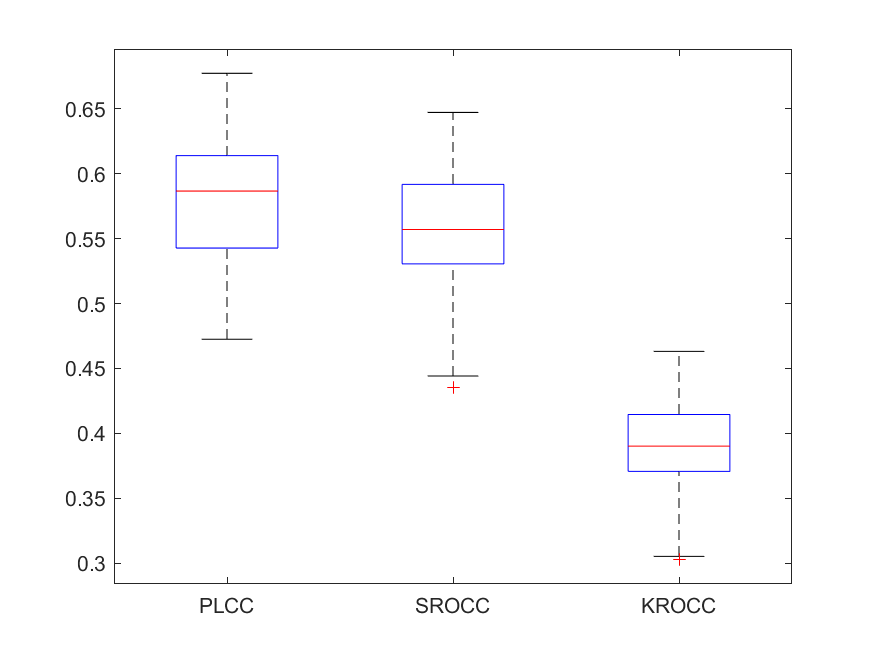}
        \caption{GRAD-LOG-CP.}
    \end{subfigure}
    ~ 
    \begin{subfigure}[b]{0.45\textwidth}
        \includegraphics[width=\textwidth]{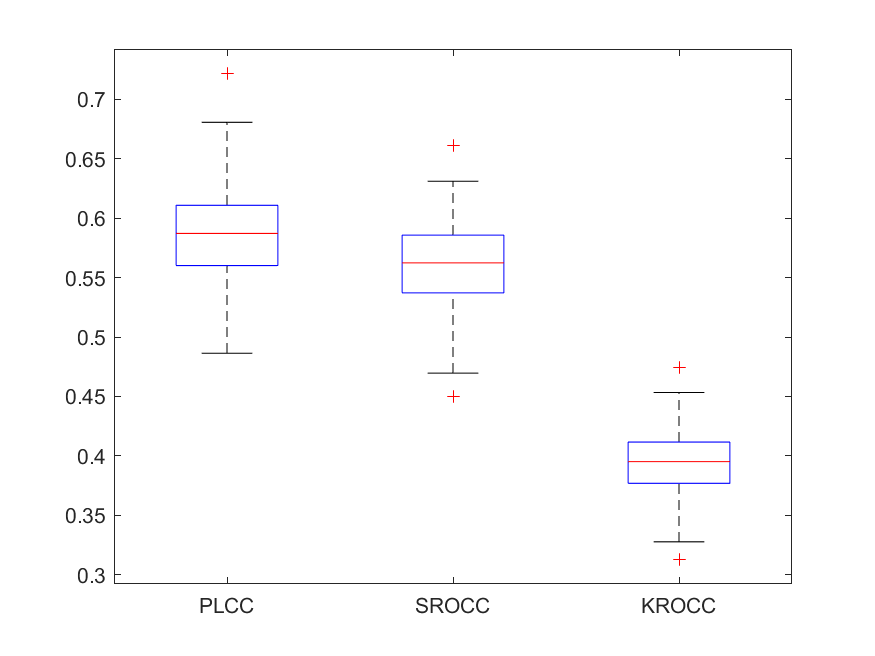}
        \caption{GWH-GLBP-BIQA.}
    \end{subfigure}

    \quad
    
    \begin{subfigure}[b]{0.45\textwidth}
        \includegraphics[width=\textwidth]{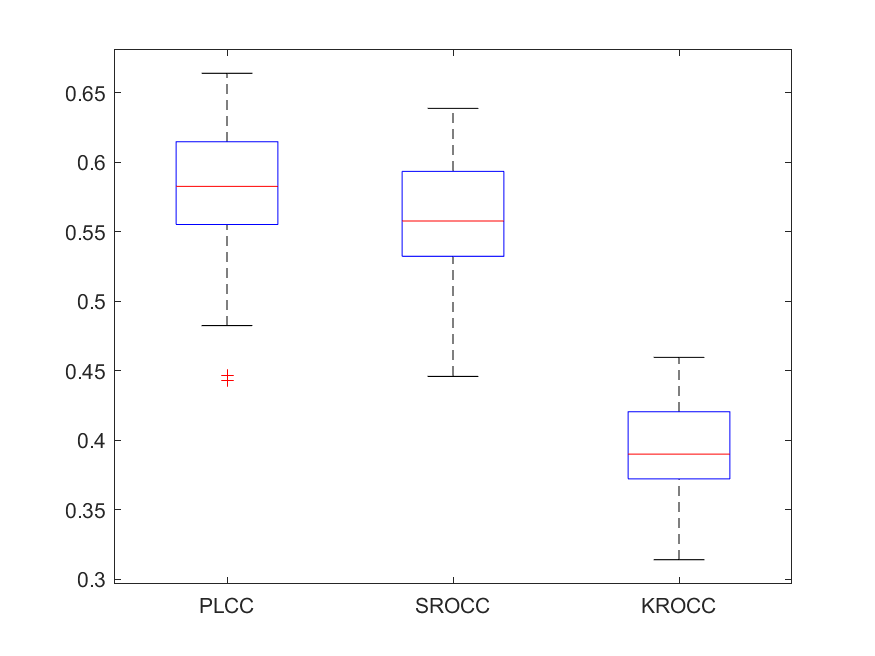}
        \caption{IQVG.}
    \end{subfigure}
    ~
    \begin{subfigure}[b]{0.45\textwidth}
        \includegraphics[width=\textwidth]{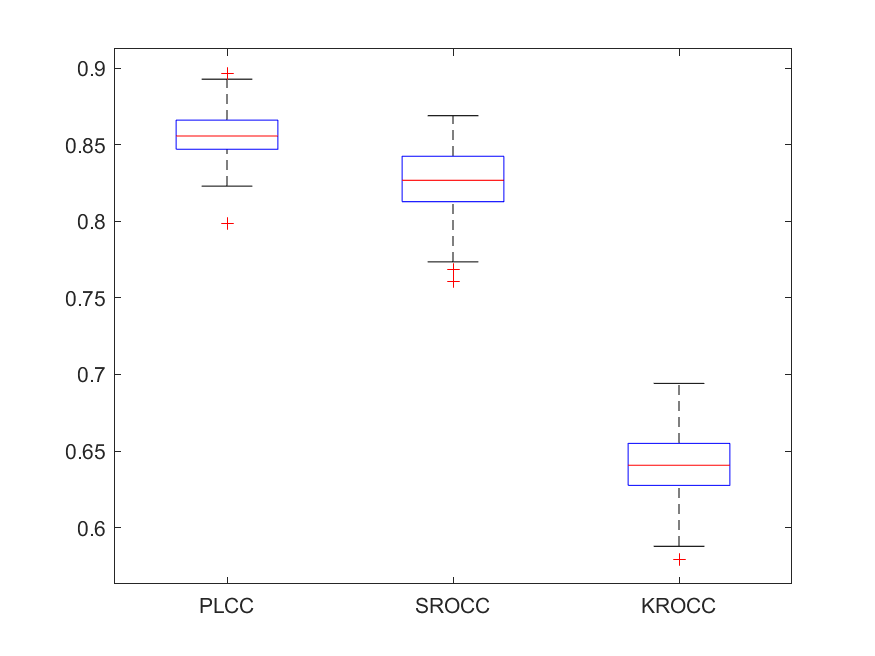}
        \caption{MultiGAP-GPR.}
    \end{subfigure}

    \quad

    \begin{subfigure}[b]{0.45\textwidth}
        \includegraphics[width=\textwidth]{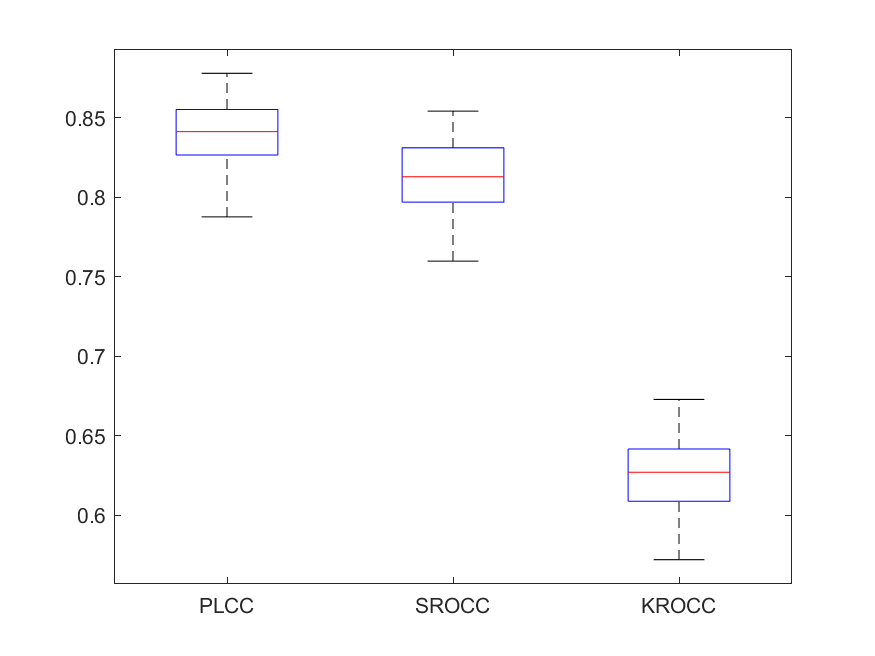}
        \caption{MultiGAP-SVR.}
    \end{subfigure}
    ~
    \begin{subfigure}[b]{0.45\textwidth}
        \includegraphics[width=\textwidth]{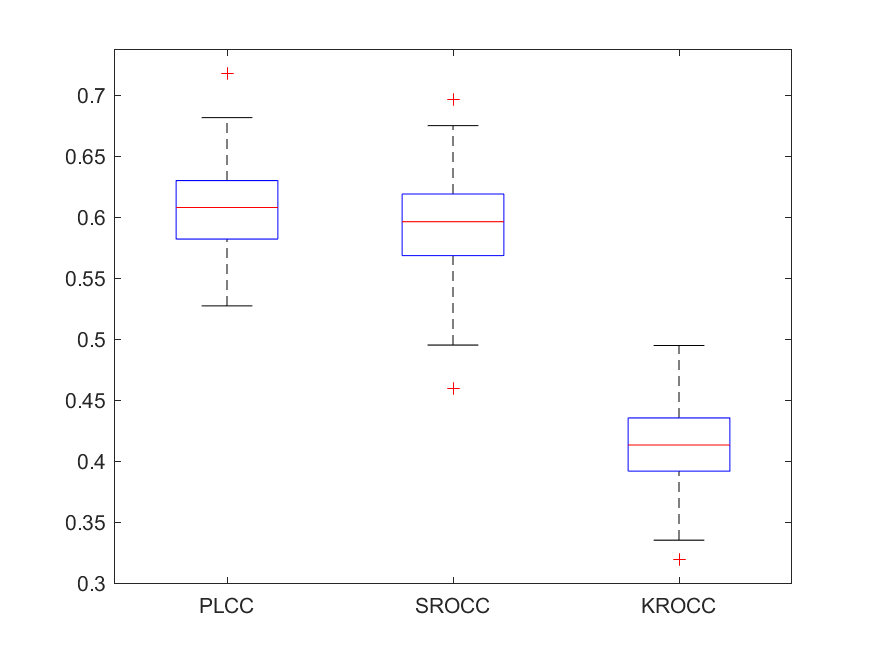}
        \caption{NBIQA.}
    \end{subfigure}
    \caption{Box plots of the measured PLCC, SROCC, and KROCC values on LIVE In the Wild database.
    100 random train-test splits.}
    \label{fig:boxplot2}
\end{figure}

\begin{figure}
    \centering
    \begin{subfigure}[b]{0.45\textwidth}
        \includegraphics[width=\textwidth]{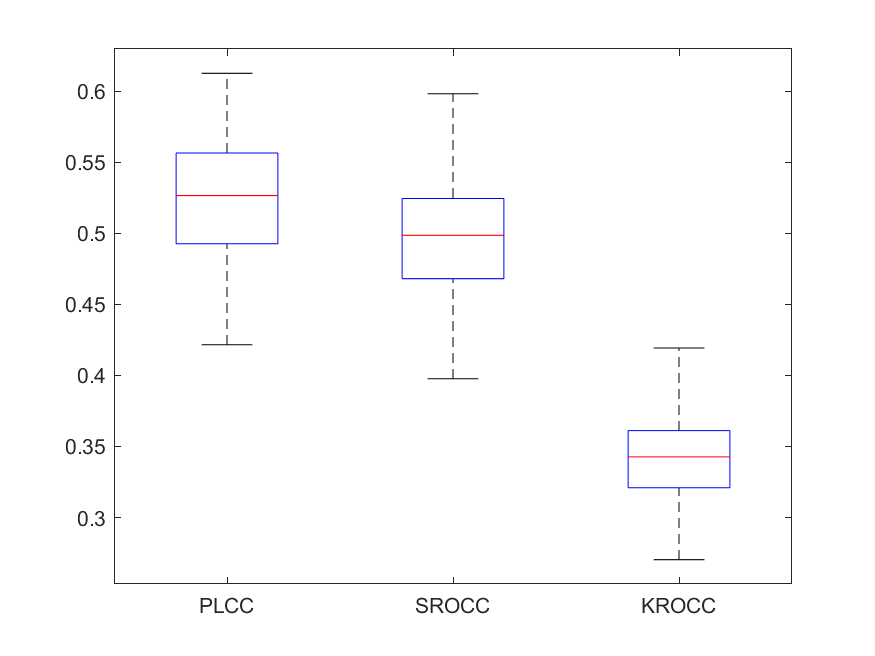}
        \caption{OG-IQA.}
    \end{subfigure}
    ~ 
    \begin{subfigure}[b]{0.45\textwidth}
        \includegraphics[width=\textwidth]{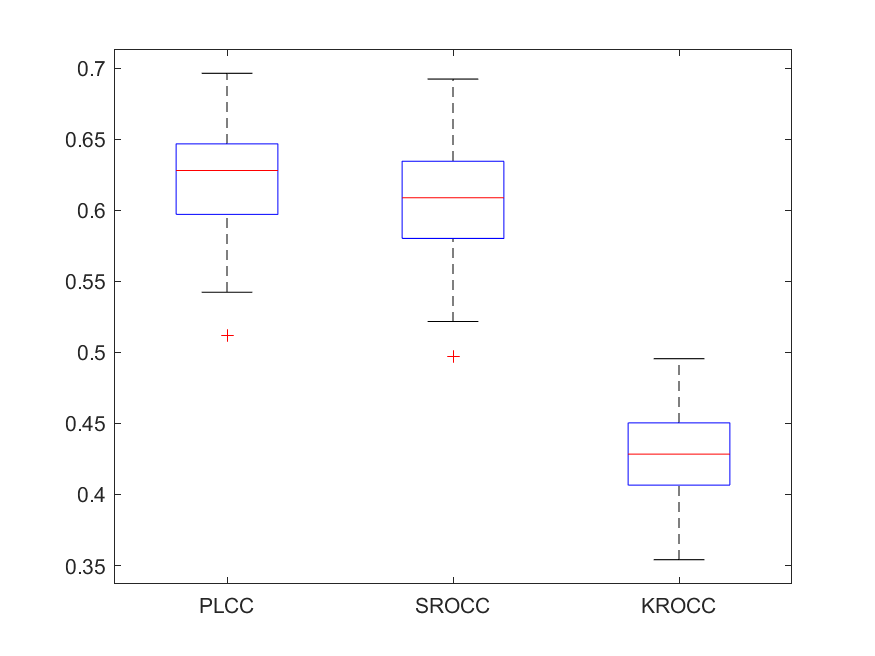}
        \caption{ORACLE.}
    \end{subfigure}

    \quad
    
    \begin{subfigure}[b]{0.45\textwidth}
        \includegraphics[width=\textwidth]{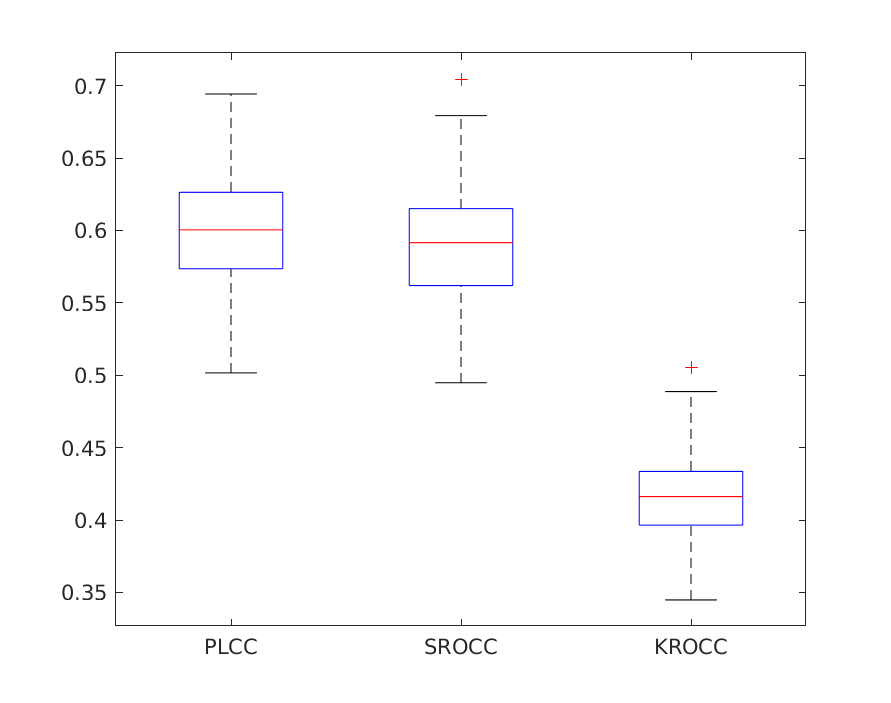}
        \caption{SCORER.}
    \end{subfigure}
    ~
    \begin{subfigure}[b]{0.45\textwidth}
        \includegraphics[width=\textwidth]{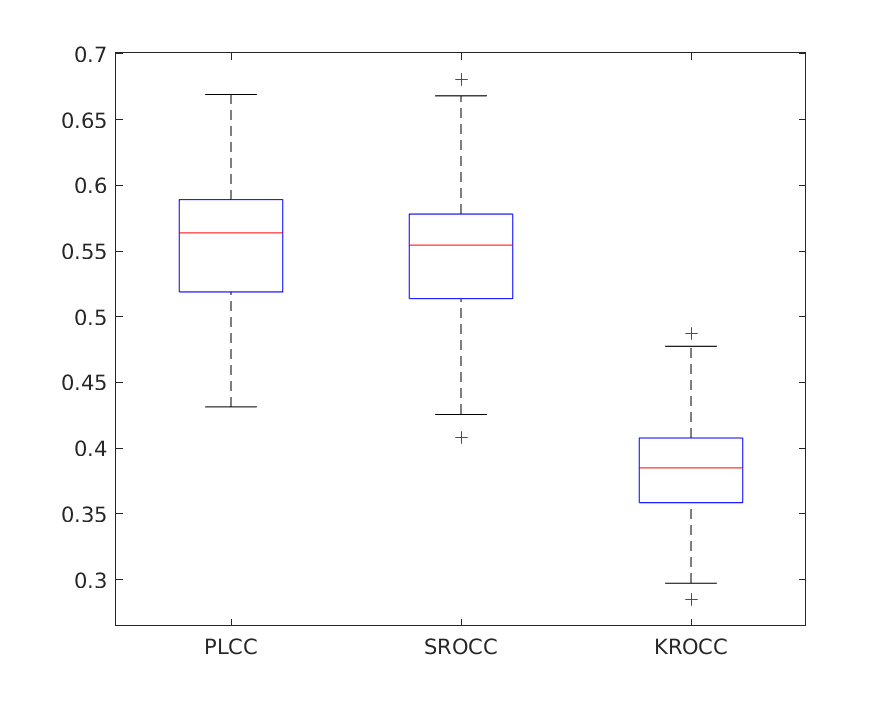}
        \caption{SEER.}
    \end{subfigure}

    \quad

    \begin{subfigure}[b]{0.45\textwidth}
        \includegraphics[width=\textwidth]{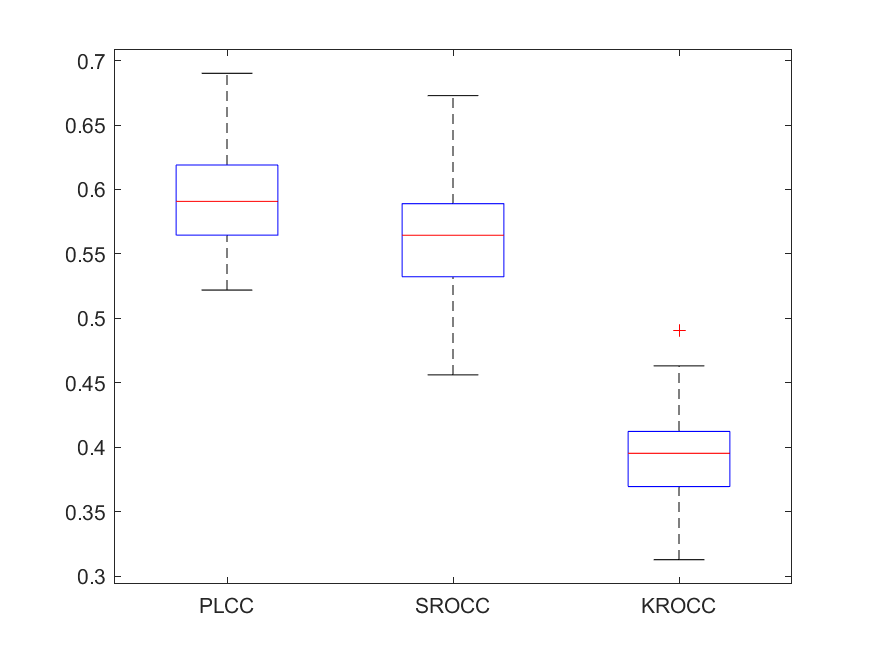}
        \caption{SPF-IQA.}
    \end{subfigure}
    ~
    \begin{subfigure}[b]{0.45\textwidth}
        \includegraphics[width=\textwidth]{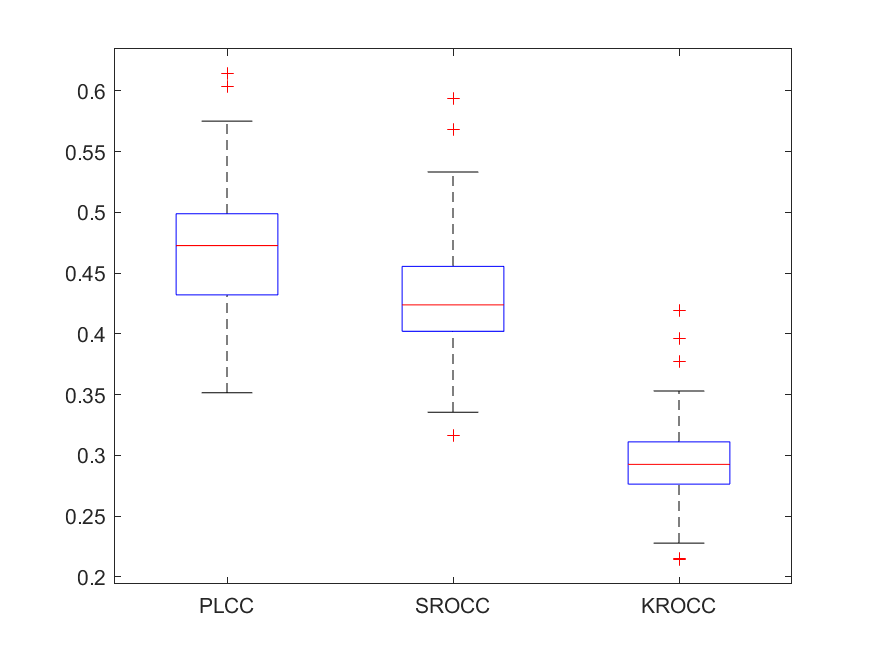}
        \caption{SSEQ.}
    \end{subfigure}
    \caption{Box plots of the measured PLCC, SROCC, and KROCC values on LIVE In the Wild database.
    100 random train-test splits.}
    \label{fig:boxplot3}
\end{figure}

In this subsection, we present our results measured on LIVE In the Wild database.
Average PLCC, SROCC, and KROCC are summarized in Table \ref{table:clive}. It can be seen
that MultiGAP \cite{varga2020multi} methods, which extracts deep features from Inception modules of an Inception-v3
convolutional neural network, achieve the best results. MultiGAP-SVR \cite{varga2020multi} 
utilizes an support vector 
regressor (SVR)
with Gaussian kernel function, while MultiGAP-GPR \cite{varga2020multi}, \cite{varga2020comprehensive} relies
on Gaussian process regression with
rational quadratic kernel function. The statistics of PLCC, SROCC, and KROCC values
are summarized in Figures \ref{fig:boxplot1}, \ref{fig:boxplot2}, and \ref{fig:boxplot3}.
On each box, the central mark indicates the median, and the bottom and
top edges of the box indicate the 25$th$ and 75$th$ percentiles, respectively.
The whiskers extend to the most extreme data points not considered outliers,
and the outliers are plotted individually using the $'+'$ symbol.

\begin{table*}[ht]
\caption{Overall performance on LIVE In the Wild. Average PLCC, SROCC, and KROCC
are reported, measured 100 random train-test splits. The best results are typed
by \textbf{bold}, the second best results are typed by \textit{italic}, the third best
results are \underline{underlined}.
} 
\centering 
\begin{center}
    \begin{tabular}{ |c|c|c|c|}
    \hline
Method&PLCC&SROCC&KROCC\\
    \hline
BLIINDS-II \cite{saad2012blind} & 0.450& 0.419& 0.292\\
BMPRI \cite{min2018blind}       & 0.521& 0.480& 0.332\\
BRISQUE \cite{mittal2012no}     & 0.503& 0.487& 0.333\\
CurveletQA \cite{liu2014no}     & 0.620& \underline{0.611}& \underline{0.433}\\
DIIVINE \cite{moorthy2011blind} & 0.602& 0.579& 0.405\\
ENIQA \cite{chen2019no}         & 0.578& 0.554& 0.386\\
GRAD-LOG-CP \cite{xue2014blind} & 0.579& 0.557& 0.391\\
GWH-GLBP-BIQA \cite{li2016no}   & 0.588& 0.560& 0.394\\
IQVG \cite{gu2013learning}      & 0.581& 0.559& 0.393\\
MultiGAP-SVR \cite{varga2020multi} & \textit{0.841}& \textit{0.813}& \textit{0.626}\\
MultiGAP-GPR \cite{varga2020multi}. \cite{varga2020comprehensive} & \textbf{0.857}& \textbf{0.826}& \textbf{0.641}\\
NBIQA \cite{ou2019novel}        & 0.607& 0.593& 0.414\\
OG-IQA \cite{liu2016blind}      & 0.526& 0.497& 0.342\\
ORACLE \cite{oszust2018optimized} & \underline{0.622}& 0.606& 0.427\\
PIQE \cite{venkatanath2015blind}& 0.171& 0.108& 0.070\\
SCORER \cite{oszust2019local}   & 0.599& 0.590& 0.416\\
SEER \cite{oszust2019no}        & 0.556& 0.547& 0.383\\
SPF-IQA \cite{varga2020no}      & 0.592& 0.563& 0.395\\
SSEQ \cite{liu2014noS}          & 0.469& 0.429& 0.295\\
 \hline
 \end{tabular}
\end{center}
\label{table:clive}
\end{table*}

\subsection{KonIQ-10k}
\begin{figure}
    \centering
    \begin{subfigure}[b]{0.45\textwidth}
        \includegraphics[width=\textwidth]{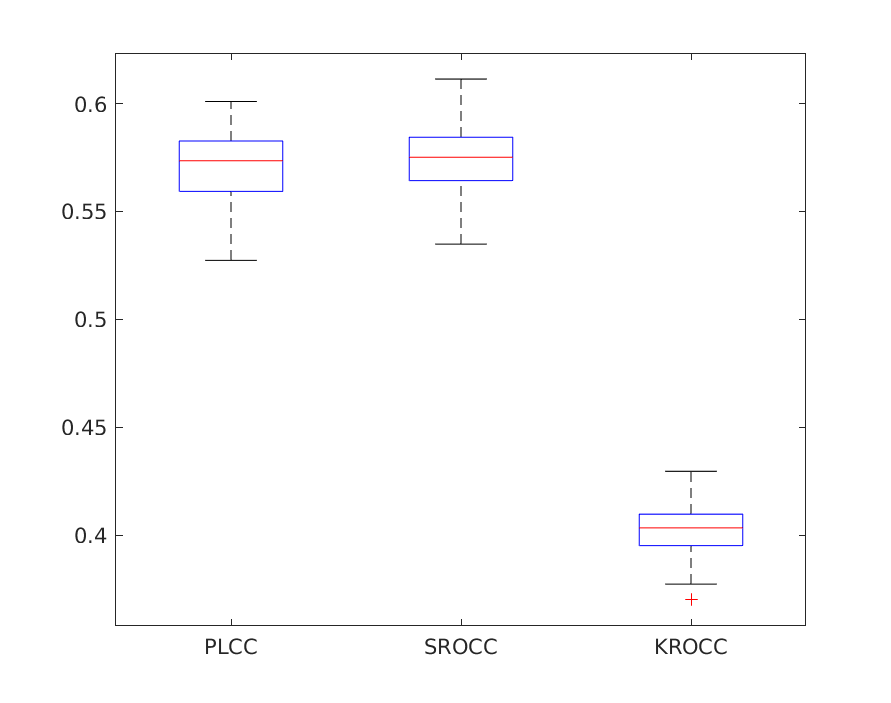}
        \caption{BLIINDS-II.}
    \end{subfigure}
    ~ 
    \begin{subfigure}[b]{0.45\textwidth}
        \includegraphics[width=\textwidth]{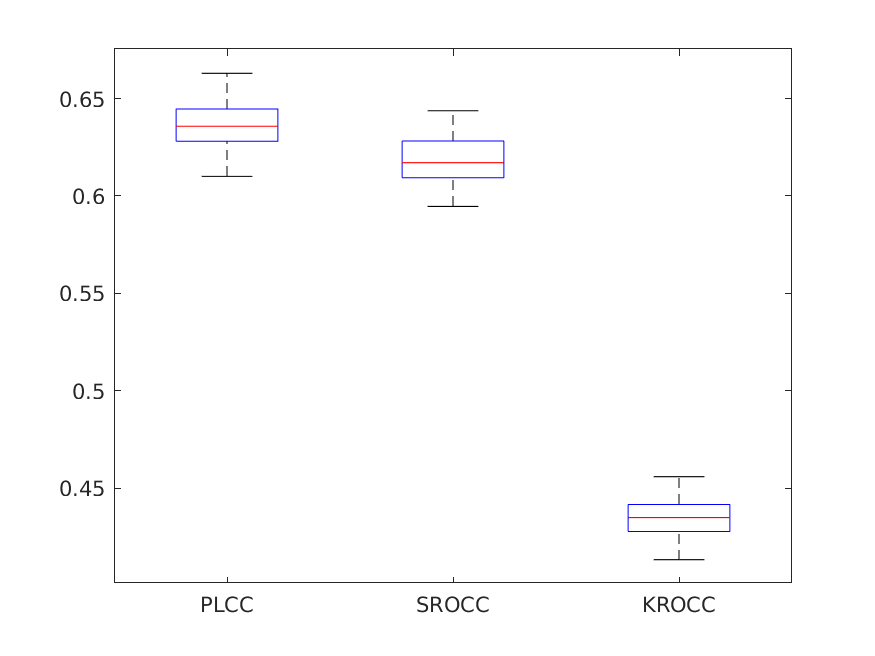}
        \caption{BMPRI.}
    \end{subfigure}

    \quad
    
    \begin{subfigure}[b]{0.45\textwidth}
        \includegraphics[width=\textwidth]{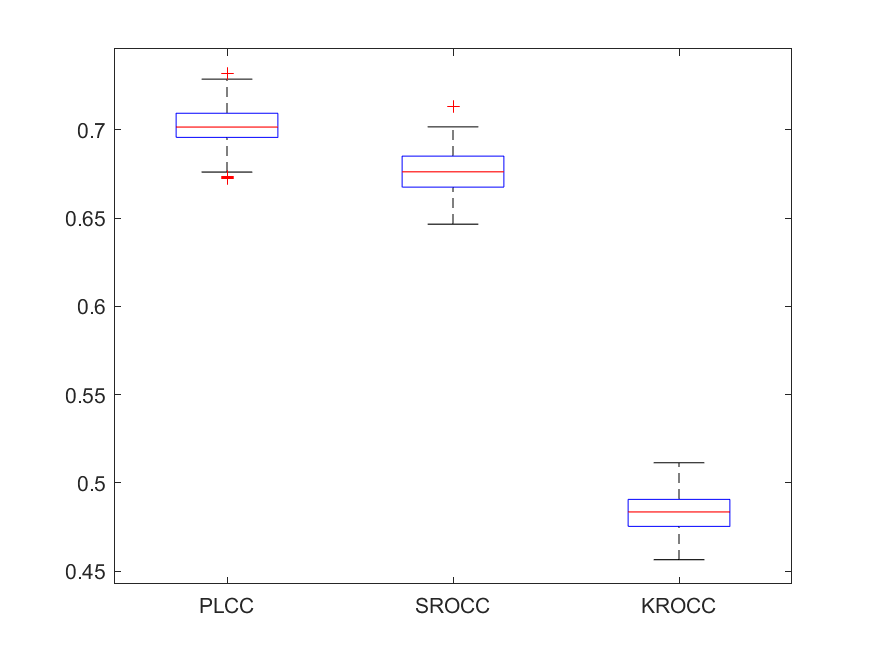}
        \caption{BRISQUE.}
    \end{subfigure}
    ~
    \begin{subfigure}[b]{0.45\textwidth}
        \includegraphics[width=\textwidth]{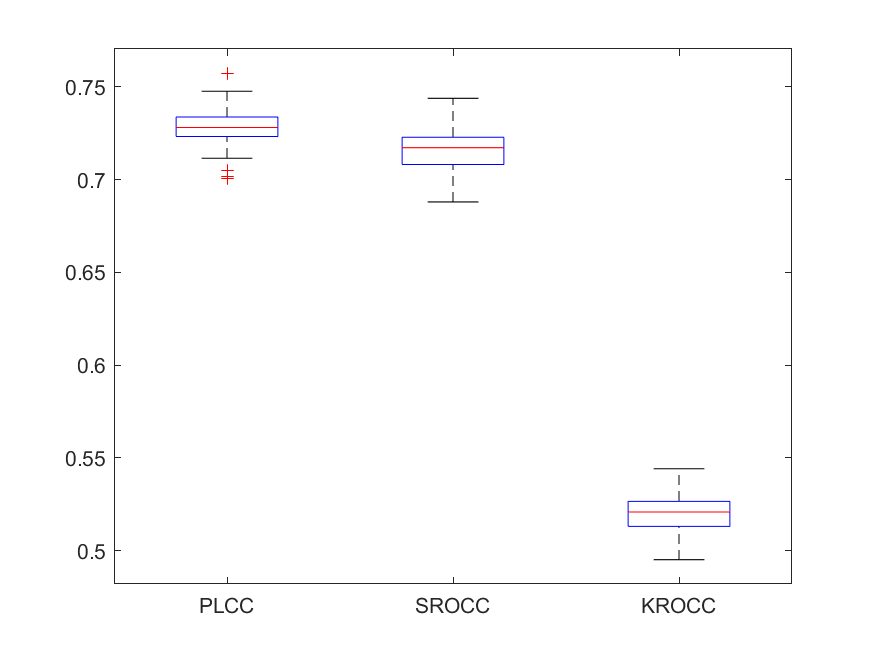}
        \caption{CurveletQA.}
    \end{subfigure}

    \quad

    \begin{subfigure}[b]{0.45\textwidth}
        \includegraphics[width=\textwidth]{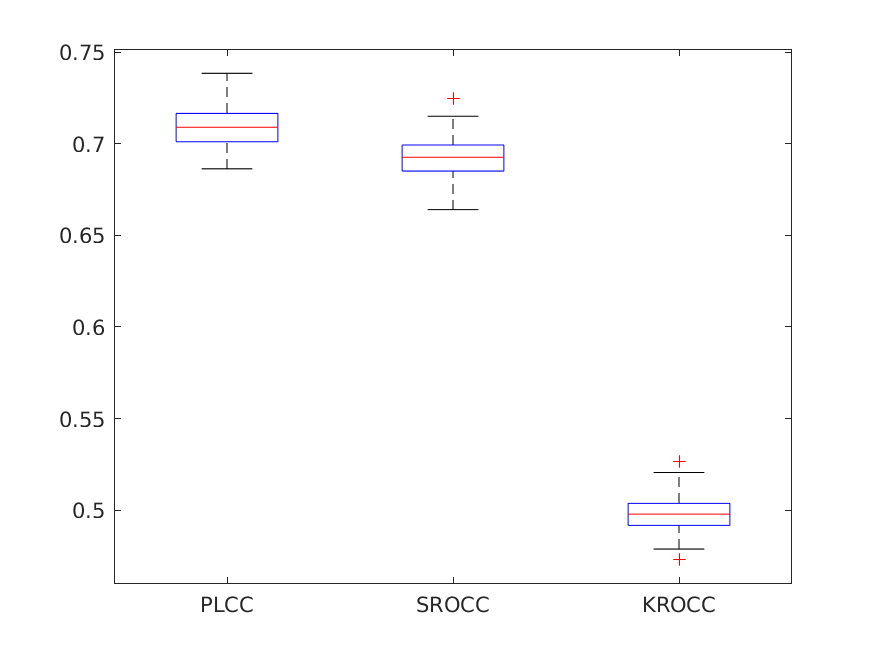}
        \caption{DIIVINE.}
    \end{subfigure}
    ~
    \begin{subfigure}[b]{0.45\textwidth}
        \includegraphics[width=\textwidth]{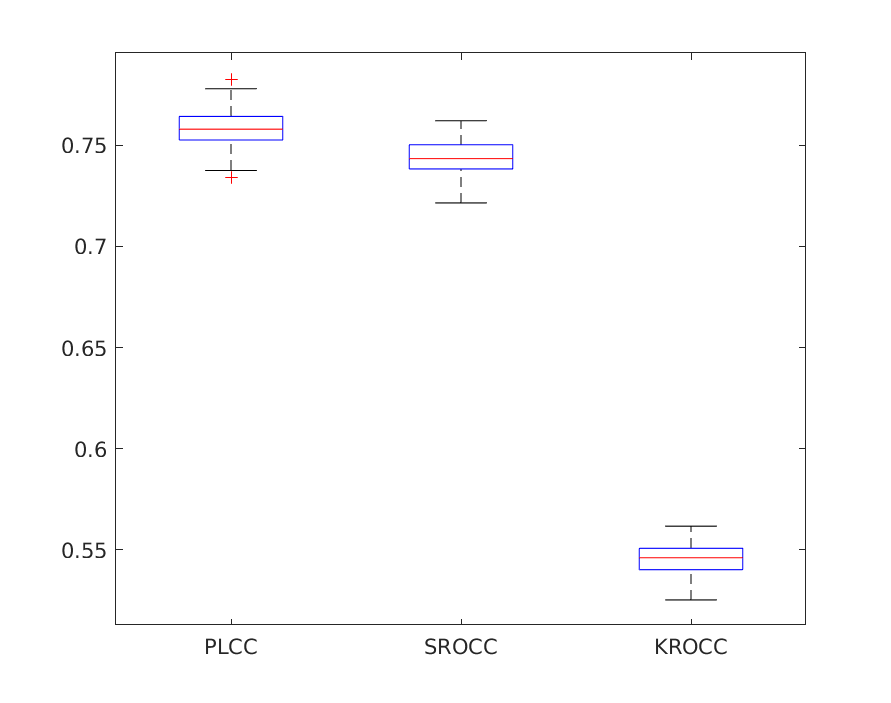}
        \caption{ENIQA.}
    \end{subfigure}
    \caption{Box plots of the measured PLCC, SROCC, and KROCC values on KonIQ-10k database.
    100 random train-test splits.}
    \label{fig:boxplot4}
\end{figure}

\begin{figure}
    \centering
    \begin{subfigure}[b]{0.45\textwidth}
        \includegraphics[width=\textwidth]{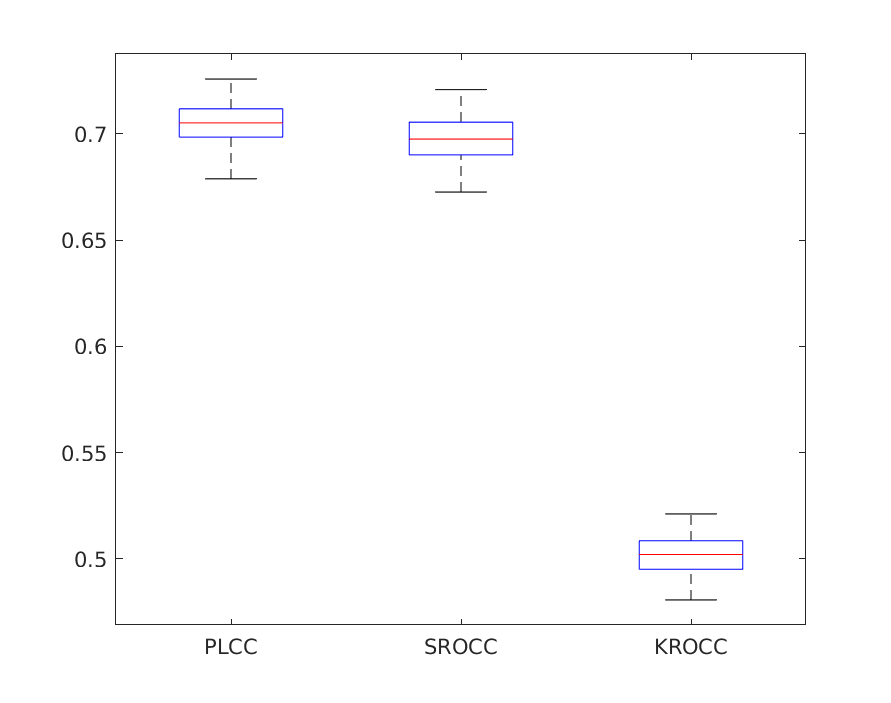}
        \caption{GRAD-LOG-CP.}
    \end{subfigure}
    ~ 
    \begin{subfigure}[b]{0.45\textwidth}
        \includegraphics[width=\textwidth]{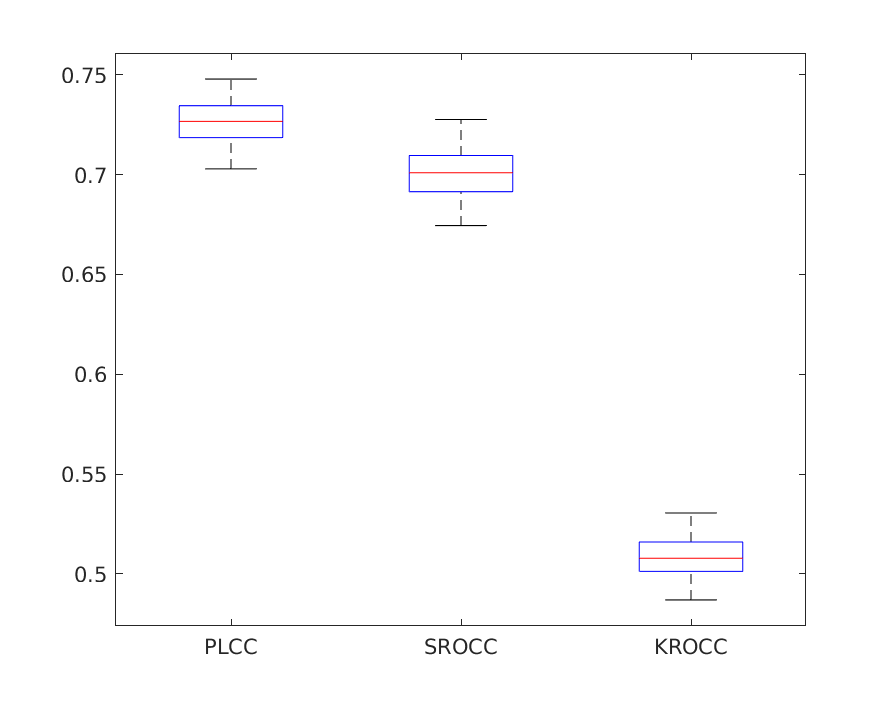}
        \caption{GWH-GLBP-BIQA.}
    \end{subfigure}

    \quad
    
    \begin{subfigure}[b]{0.45\textwidth}
        \includegraphics[width=\textwidth]{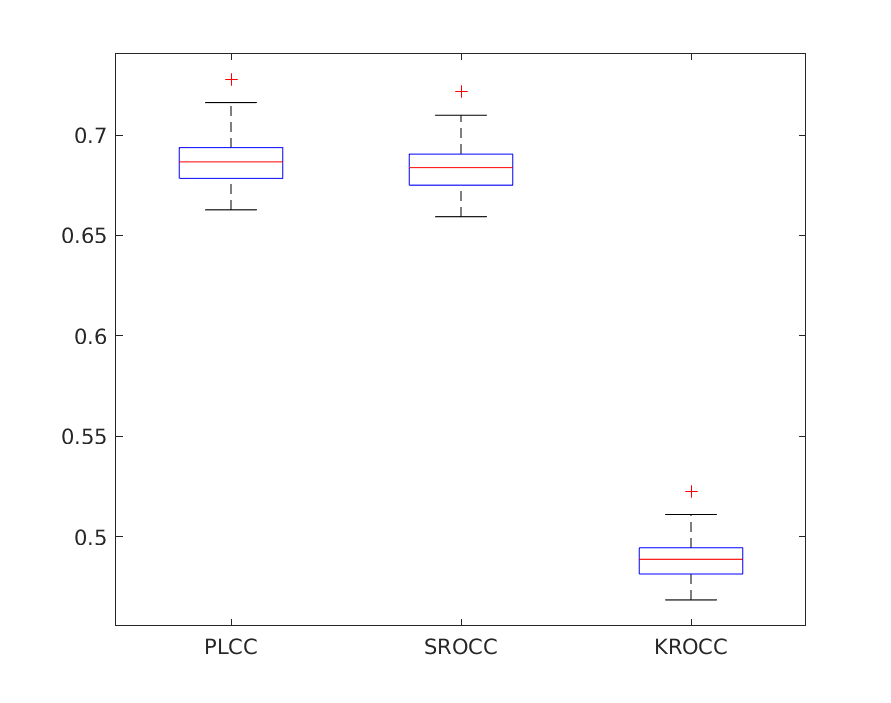}
        \caption{IQVG.}
    \end{subfigure}
    ~
    \begin{subfigure}[b]{0.45\textwidth}
        \includegraphics[width=\textwidth]{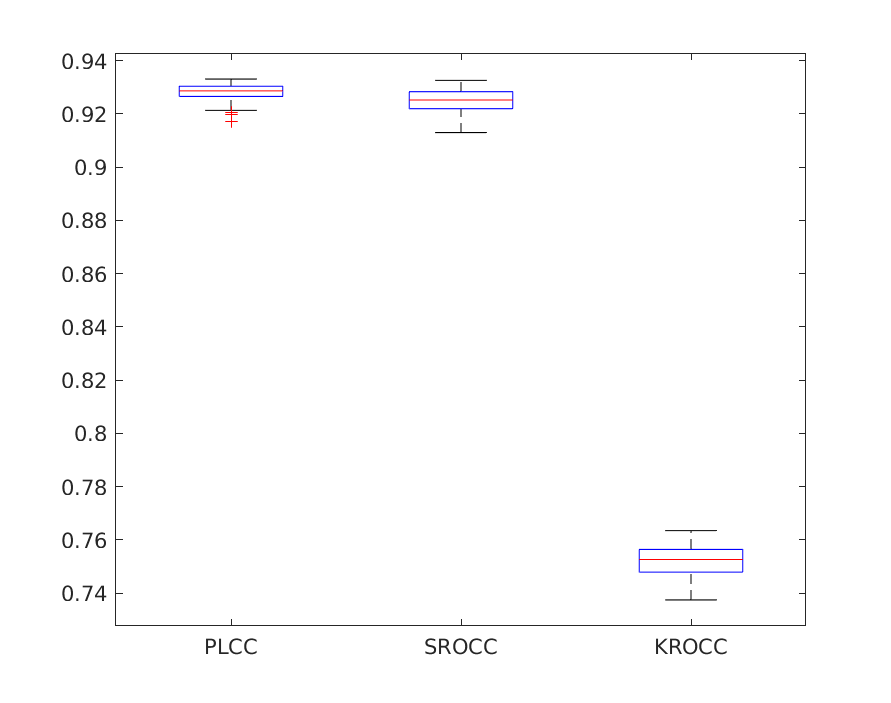}
        \caption{MultiGAP-GPR.}
    \end{subfigure}

    \quad

    \begin{subfigure}[b]{0.45\textwidth}
        \includegraphics[width=\textwidth]{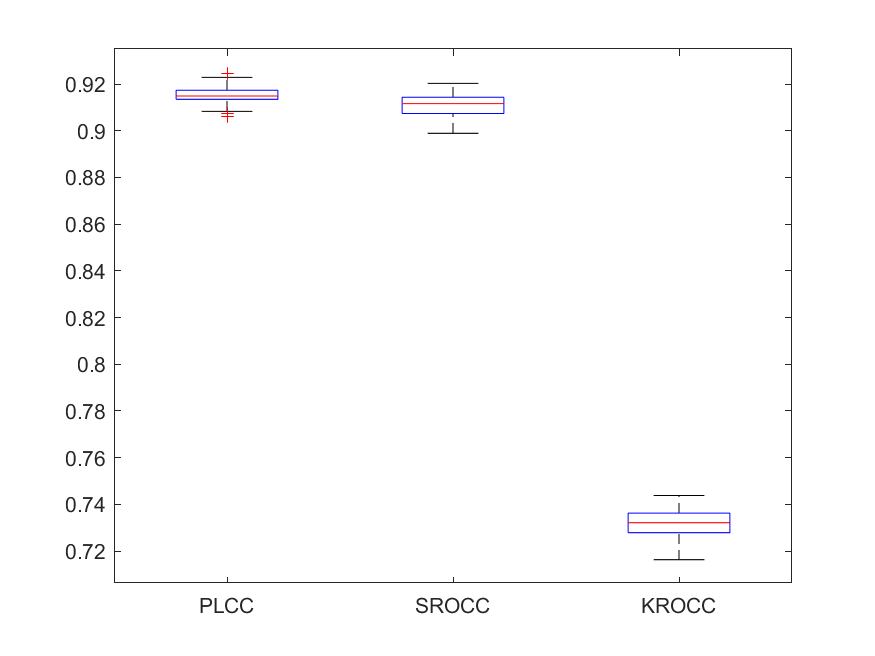}
        \caption{MultiGAP-SVR.}
    \end{subfigure}
    ~
    \begin{subfigure}[b]{0.45\textwidth}
        \includegraphics[width=\textwidth]{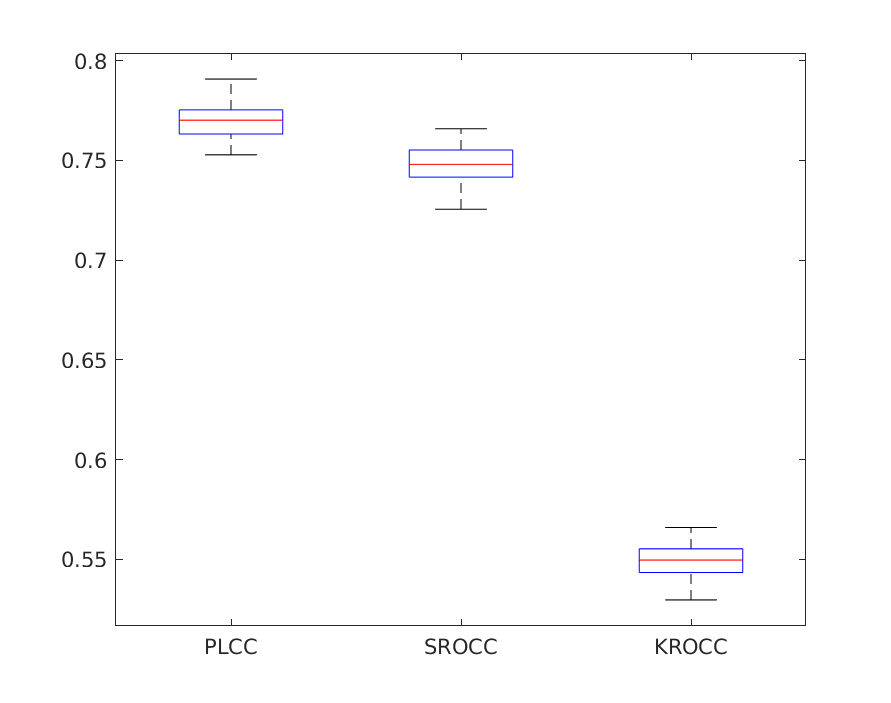}
        \caption{NBIQA.}
    \end{subfigure}
    \caption{Box plots of the measured PLCC, SROCC, and KROCC values on KonIQ-10k database.
    100 random train-test splits.}
    \label{fig:boxplot5}
\end{figure}

\begin{figure}
    \centering
    \begin{subfigure}[b]{0.45\textwidth}
        \includegraphics[width=\textwidth]{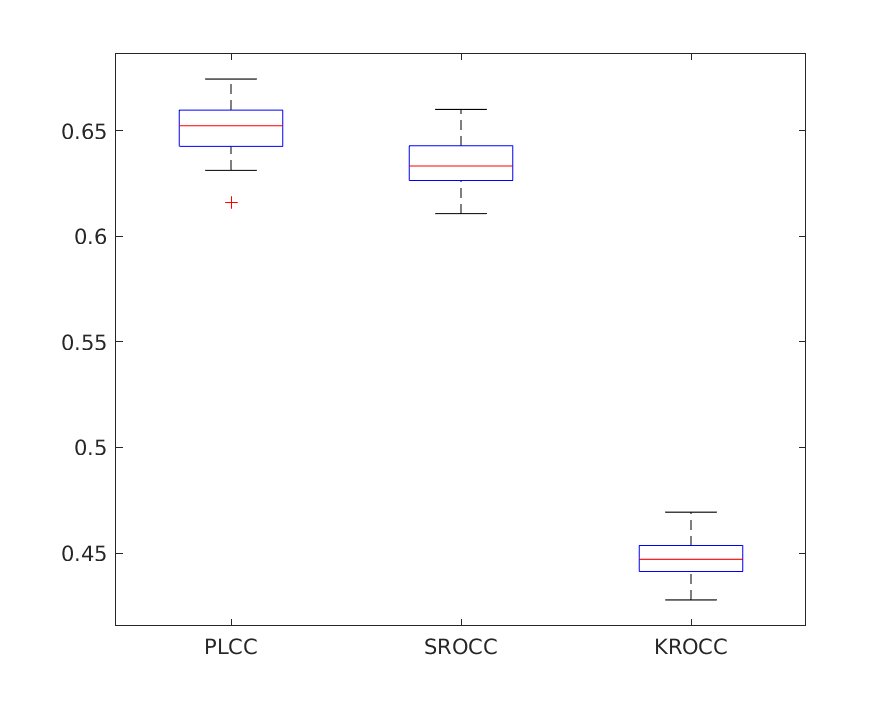}
        \caption{OG-IQA.}
    \end{subfigure}
    ~ 
    \begin{subfigure}[b]{0.45\textwidth}
        \includegraphics[width=\textwidth]{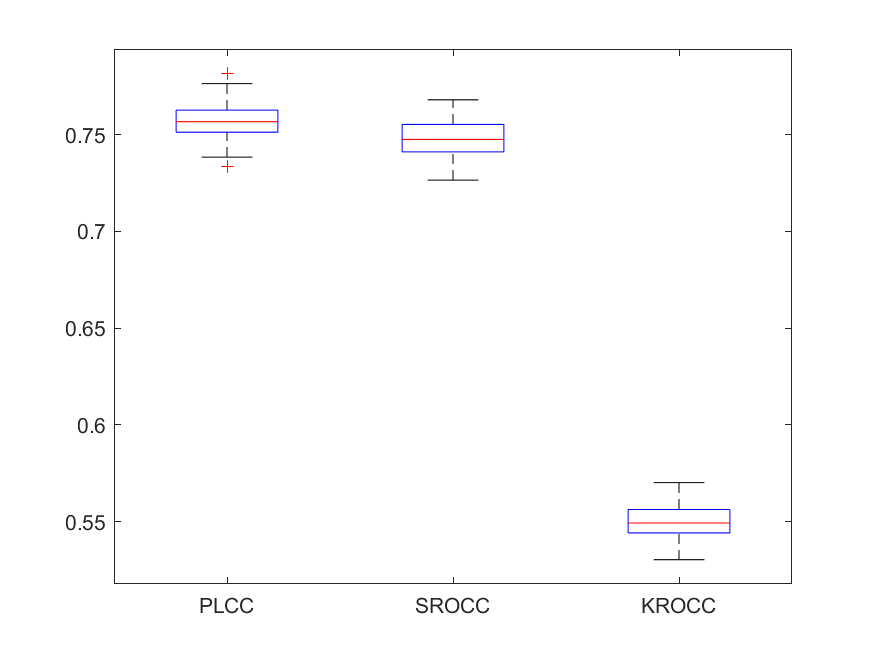}
        \caption{ORACLE.}
    \end{subfigure}

    \quad
    
    \begin{subfigure}[b]{0.45\textwidth}
        \includegraphics[width=\textwidth]{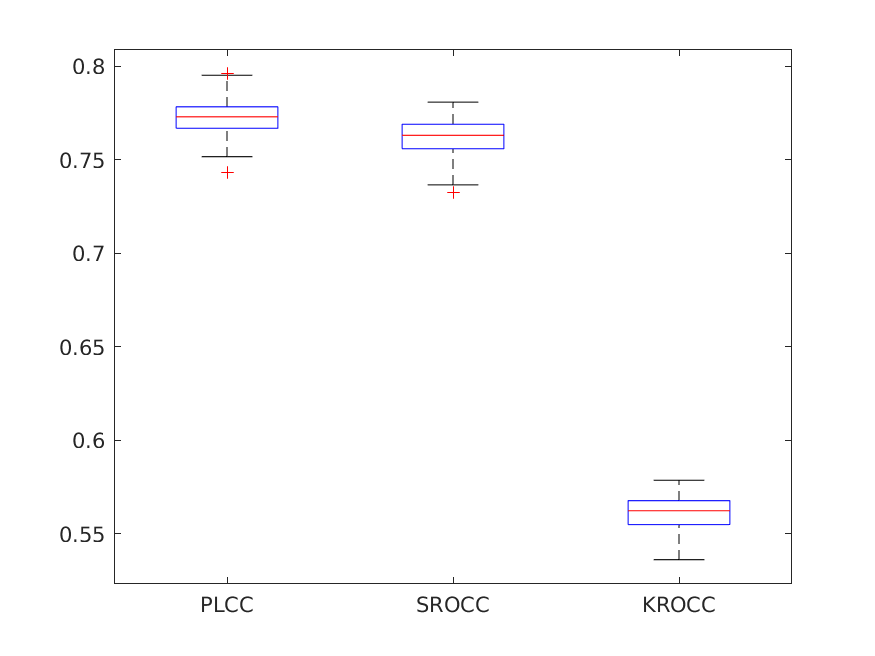}
        \caption{SCORER.}
    \end{subfigure}
    ~
    \begin{subfigure}[b]{0.45\textwidth}
        \includegraphics[width=\textwidth]{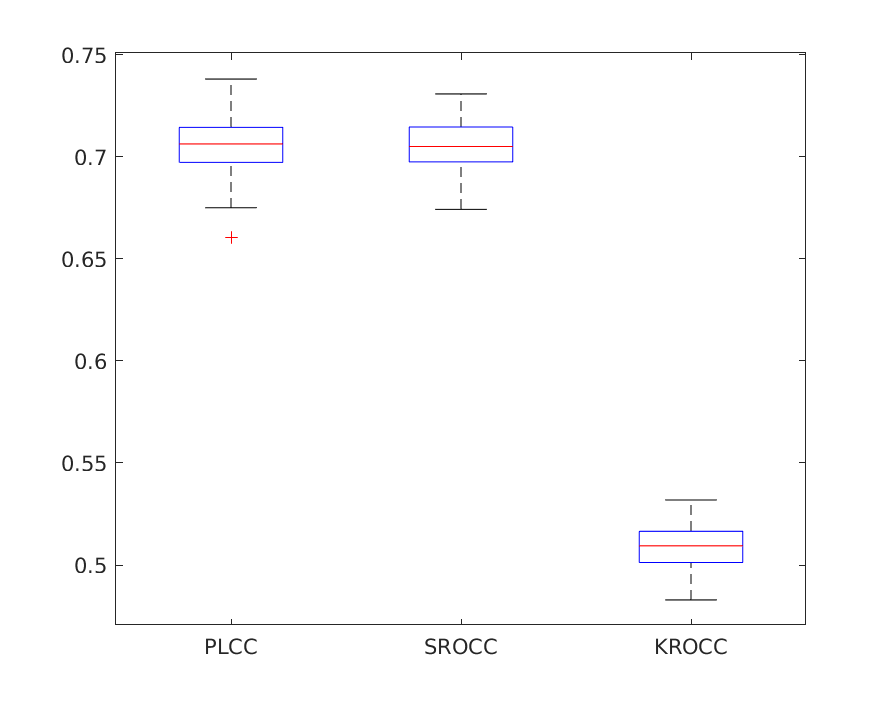}
        \caption{SEER.}
    \end{subfigure}

    \quad

    \begin{subfigure}[b]{0.45\textwidth}
        \includegraphics[width=\textwidth]{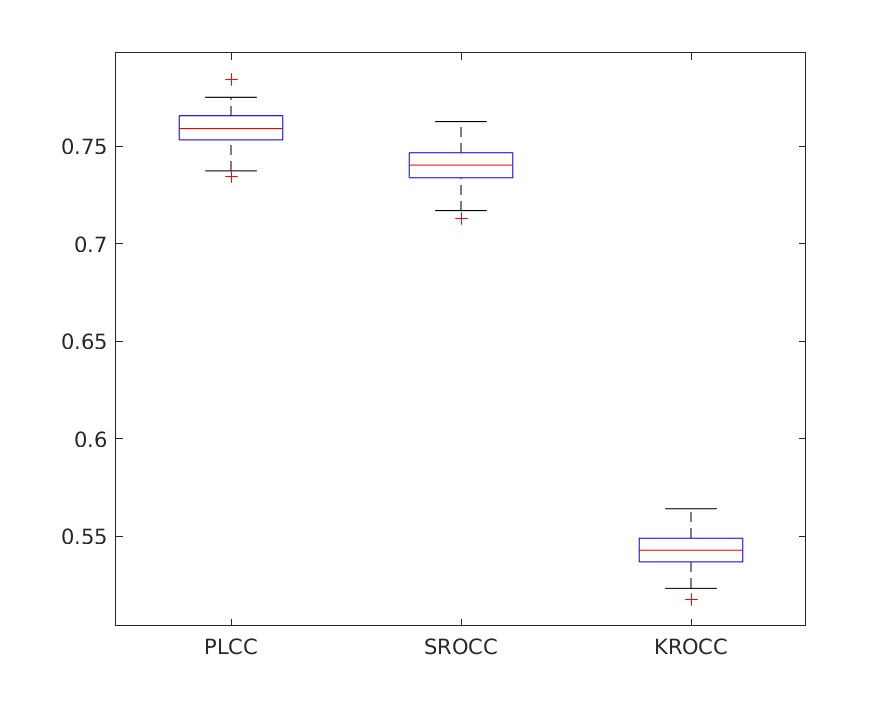}
        \caption{SPF-IQA.}
    \end{subfigure}
    ~
    \begin{subfigure}[b]{0.45\textwidth}
        \includegraphics[width=\textwidth]{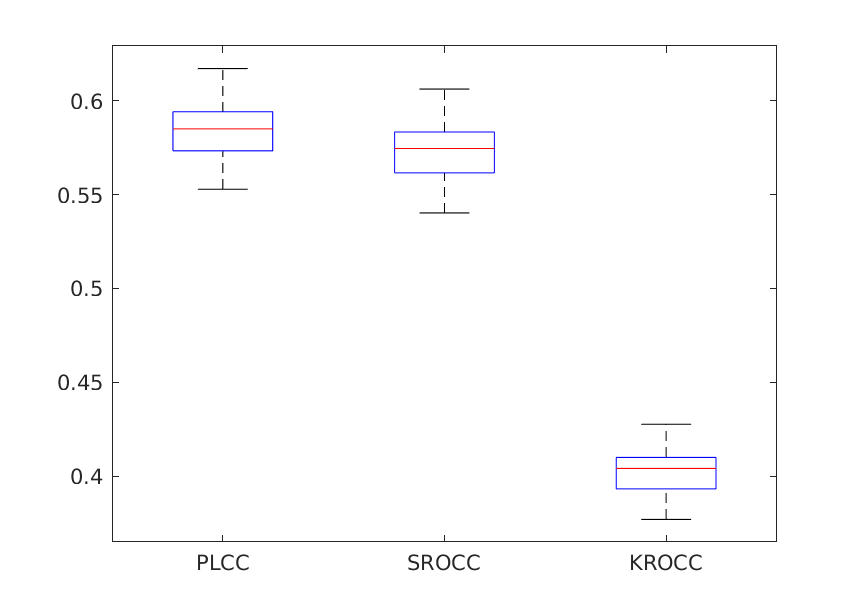}
        \caption{SSEQ.}
    \end{subfigure}
    \caption{Box plots of the measured PLCC, SROCC, and KROCC values on KonIQ-10k database.
    100 random train-test splits.}
    \label{fig:boxplot6}
\end{figure}

In this subsection, we present our results measured on KonIQ-10k database. Average PLCC, SROCC, and KROCC
are summarized in Table \ref{table:koniq}. Similar to LIVE In the Wild,
MultiGAP \cite{varga2020multi}, \cite{varga2020comprehensive} methods achieve the best
performance. The statistics of PLCC, SROCC, and KROCC values are summarized in Figures
\ref{fig:boxplot4}, \ref{fig:boxplot5}, and \ref{fig:boxplot6}. As already mentioned, 
on each box, the central mark indicates the median, and the bottom and
top edges of the box indicate the 25$th$ and 75$th$ percentiles, respectively.
The whiskers extend to the most extreme data points not considered outliers,
and the outliers are plotted individually using the $'+'$ symbol.
\begin{table*}[ht]
\caption{Overall performance on KonIQ-10k. Average PLCC, SROCC, and KROCC
are reported, measured 100 random train-test splits. The best results are typed
by \textbf{bold}, the second best results are typed by \textit{italic}, the third best
results are \underline{underlined}.
} 
\centering 
\begin{center}
    \begin{tabular}{ |c|c|c|c|}
    \hline
Method&PLCC&SROCC&KROCC\\
    \hline
BLIINDS-II \cite{saad2012blind} & 0.571 & 0.575 & 0.403\\
BMPRI \cite{min2018blind}       & 0.636& 0.619& 0.435\\
BRISQUE \cite{mittal2012no}     & 0.702& 0.676& 0.483\\
CurveletQA \cite{liu2014no}     & 0.728& 0.716& 0.520\\
DIIVINE \cite{moorthy2011blind} & 0.709& 0.692& 0.498\\
ENIQA \cite{chen2019no}         & 0.758& 0.744& 0.546\\
GRAD-LOG-CP \cite{xue2014blind} & 0.705& 0.698& 0.502\\
GWH-GLBP-BIQA \cite{li2016no}    & 0.726& 0.700& 0.508\\
IQVG \cite{gu2013learning}       & 0.688 & 0.684 & 0.489\\
MultiGAP-SVR \cite{varga2020multi} & \textit{0.915}& \textit{0.911}& \textit{0.732}\\
MultiGAP-GPR \cite{varga2020multi}, \cite{varga2020comprehensive}&\textbf{0.928}&\textbf{0.925}&\textbf{0.752}\\
NBIQA \cite{ou2019novel}          & 0.770& 0.748& 0.549\\
OG-IQA \cite{liu2016blind}         & 0.652& 0.635& 0.448\\
ORACLE \cite{oszust2018optimized} & 0.757& 0.748& 0.550\\
PIQE \cite{venkatanath2015blind}& 0.206& 0.245& 0.165\\
SCORER \cite{oszust2019local}   & \underline{0.772}& \underline{0.762}& \underline{0.561}\\
SEER \cite{oszust2019no}        & 0.705& 0.705& 0.509\\
SPF-IQA \cite{varga2020no}      & 0.759& 0.740& 0.543\\
SSEQ \cite{liu2014noS}          & 0.584& 0.573& 0.402\\
 \hline
 \end{tabular}
\end{center}
\label{table:koniq}
\end{table*}

\section{Conclusion}
\label{sec:conclusion}
In this study, we evaluated several NR-IQA algorithms on authentic distortions. Specifically, we utilized
two publicly available benchmark databases (LIVE In the Wild and KonIQ-10k) to evaluate the state-of-the-art
in NR-IQA. Appx. 80$\%$ of images were used for training and the remaining 20$\%$ were used for testing.
Average PLCC, SROCC, and KROCC values were reported measured over 100 random train-test splits.
Moreover, the boxplots of PLCC, SROCC, and KROCC values were also published for each examined machine learning
based method. Our evaluation results may be helpful to obtain a clear
understanding about the status of state-of-the-art no-reference image quality assessment methods.
\bibliographystyle{unsrt}  

\bibliography{references}

\end{document}